\PassOptionsToPackage{super}{natbib}
\documentclass[twocolumn,english,2pt]{revtex4-1}
\usepackage{amsmath}
\usepackage{amssymb}
\usepackage{cancel}
\usepackage{graphicx}
\usepackage{esint}
\usepackage{tabularx}
\usepackage[super]{natbib}
\setcitestyle{authoryear,open={},close={},super}
\makeatletter
\usepackage{epstopdf}
\usepackage{float}
\usepackage{textcomp,booktabs}
\usepackage[usenames,dvipsnames]{color}
\usepackage{colortbl}
\definecolor{mygray}{gray}{.9}
\definecolor{mypink}{rgb}{.99,.91,.95}
\definecolor{mycyan}{cmyk}{.3,0,0,0}
\definecolor{myblue}{rgb}{1,0.9,0.6}
\usepackage{dcolumn}
\usepackage{bm}
\usepackage{enumerate}
\usepackage{float}
\usepackage{calc}

\usepackage{babel}
\addto\captionsenglish{}

\begin{document}
\title{Non-Floquet engineering in periodically driven non-Hermitian systems}
\author{Huan-Yu Wang $^1$, Xiao-Ming Zhao $^1$, Lin Zhuang $^2$ and  Wu-Ming Liu $^1$ }
\affiliation{$^1$ Beijing National Laboratory for Condensed Matter Physics, Institute of Physics, Chinese Academy of Sciences, Beijing 100190, China}
\affiliation{$^2$ School of Physics, Sun Yat-Sen University, Guangzhou 510275, People's Republic of China  }
\begin{abstract}
Floquet engineering, modulating quantum systems in a time periodic way, lies at the central part for realizing novel topological dynamical states. Thanks to the Floquet engineering, various new realms on experimentally simulating topological materials have emerged. Conventional  Floquet engineering, however, only applies to time periodic non-dissipative Hermitian systems, and for the quantum systems in reality, non-Hermitian process with dissipation usually occurs. So far, it remains unclear how to characterize topological phases of periodically driven non-Hermitian systems via the frequency space Floquet Hamiltonian. Here, we propose the non-Floquet theory to identify different Floquet topological phases of time periodic non-Hermitian systems via the generation of Floquet band gaps in frequency space. In non-Floquet theory, the eigenstates of non-Hermitian Floquet Hamiltonian are temporally deformed to be of Wannier-Stark localization. Remarkably, we show that different choices of starting points of driving period can result to different localization behavior, which effect can reversely be utilized to design detectors of quantum phases in  dissipative oscillating fields. Our protocols establish a fundamental rule for describing topological features in non-Hermitian dynamical systems and can find its applications to construct new types of Floquet topological materials.
\end{abstract}
\maketitle
Floquet engineering is widely used  in simulating various kinds of topological materials in cold atom systems\cite{RevModPhys.89.011004,NRPrUDNER, PhysRevLett.119.123601,PhysRevLett.125.183001,PhysRevX.6.041001,PhysRevLett.124.057001,PhysRevLett.111.047002}, quantum circuits\cite{PhysRevLett.122.150605,PhysRevLett.120.040404,PhysRevB.98.134204}, and  photonic systems as well\cite{NCFC,Natmaterialmac,Natphysxiao,PhysRevLett.126.113901,PhysRevLett.125.237201,Naturef, EPL,PhysRevLett.102.123905,RIP}. Topological properties are unique in time periodic systems, as the eigenenergy of the Floquet operator can not be sorted by the amplitude, but winds around with the same periodicity of driven field, which is also denoted as quasienrgy.  Consequently, apart from the zero band gap, the system may also feature a band gap pinned at quasienergy $\epsilon=\pm\frac{\pi}{T}$. The gapless edge modes centred at both $0,\pm \frac{\pi}{T}$ are anomalous, which correspond to  vanished bulk topological invariants\cite{PhysRevX.3.031005}, and the realization of Floquet topological materials with anomalous edge modes has been making increasing progress over the years.

Despite the intriguing features above, the existing framework for Floquet topological phases mainly concerns non-dissipative Hermitian processes, although the system can exchange energy and particles with the external driven field. In addition, there has been growing interest in non-Hermitian systems\cite{PhysRevLett.80.5243,IOPBender}, especially topological physics\cite{PhysRevLett.124.056802,PhysRevX.9.041015,nphycommulijm,PhysRevLett.126.086801,PhysRevLett.124.070402,PhysRevLett.122.237601,PhysRevLett.121.136802,PhysRevX.8.031079, nphys-hui,PhysRevB.103.L041115,PhysRevB.101.045415,Natphysxiao2,PNAS,PhysRevResearch.3.023022, PhysRevB.102.041119,NatCommuueda}. When a lattice system is engaged with spacial asymmetric tunneling, not only the edge modes but also the bulk states can pile up at one side, which is known as the skin effect\cite{PhysRevLett.123.170401,Natcommungong,PhysRevLett.125.126402}. Particularly, the topological properties obtained with bulk states differ dramatically from those of open boundary conditions\cite{PhysRevLett.121.086803,PhysRevB.101.121116,PhysRevLett.121.026808}. To restore the bulk-edge correspondence, the non-Bloch band theory is brought up, where the non-Hermitian eigenstates are deformed to be of standing waves, along with which comes the definition Generalized Brillouin Zone (GBZ)\cite{PhysRevLett.121.086803,PhysRevLett.123.066404}. It should be manifested that both the Bloch band theory  and  Floquet theory  stem from  ordinary differential equations with periodic potential. In practice, periodically driven systems can be mapped via the Fourier transformation to a frequency space lattice model, where a uniform external field automatically emerges due to the periodicity of Floquet states, leading to the Wannier-Stark localization, through which different Floquet topological phases can be identified from the frequency space Floquet band gap in case of high frequency driven fields\cite{PhysRevX.3.031005}.

The prerequisite for the Wannier-Stark localization above lies in that the band energy scale of the static part of the time periodic Hamiltonian shall be much smaller than the driving frequency, and is supposed to be real. Typically, the prerequisite above can only be satisfied with the conventional time periodic Hermitian system, which is broken for the non-Hermitian scenario due to the features of complex eigenenergy. Particularly, for time periodic non-Hermitian system with gain-loss process of the same amplitude as driving frequency, the complex band energy scale can not be forced to be moderate enough to realize Wannier-Stark localization. To overcome difficulties above and to restore Wannier-Stark localization make the central goal for developing non-Floquet theory.
\begin{table*}
\colorbox{mycyan}{
\begin{tabular}{lccc}
\multicolumn{4}{l}{\bf Table I: Comparison of non-Bloch band theory and non-Floquet theory } \\
\midrule
 & Transformed periodicity & Transformed states & localization behavior \\
\midrule
non-Bloch band theory & $k'=k+i\kappa$  & $\Psi(x)=\beta^{x}\psi(x), \beta=e^{-ik'}$ & Skin effect: all states localized ($\mathrm{GBZ\neq BZ}$) \\
non-Floquet theory   & $\epsilon'=\epsilon+i\dot{\Gamma}$  & $\Phi(t)=e^{-i\epsilon t+\Gamma(t)}\phi(t)$ & Wannier-Stark localization (high frequency)\\
\bottomrule
\end{tabular}
}
\end{table*}

The key part of non-Floquet theory is to transform the Floquet states in a temporal non-unitary way such that:
{\setlength\abovedisplayskip{1pt}
\setlength\belowdisplayskip{1pt}
\begin{eqnarray}
\begin{aligned}
&\Psi(t)=e^{-i\epsilon t+\Gamma (t)} \psi(t)\\
&\psi(t)=\sum_m e^{im\omega t} \tilde{\psi}(m)\\
\end{aligned}
\end{eqnarray}}where $\epsilon$ is the quasienergy, and  $\Gamma(t)$ features the modification of Floquet formalism. Through equation (1), the imaginary part of the complex band energy scale can be manipulated to recover the Wannier-Stark ladder and the emergence of real space Floquet topological edge modes can be implied from the generation of Floquet band gaps. In comparison to the non-Bloch band theory for the static system, where a spacial site dependent transformation is performed so as to restore forms of standing waves, non-Floquet engineering consider a temporal transformation on Floquet states, which aims to recover Wannier-Stark localization in frequency space.  Besides, we present that the quesienergy corresponding to the transformed Floquet operator is not identical to the original one, but differs with a shift.  Generalized qusienergy is redefined to retain discrete time translational symmetry. Our proposals benefit to predict  Floquet topological phases and design dissipative photonic topological materials.

\noindent\textbf{Theory}\\
\textbf{Dynamical pseudo-Hermiticity.} In coping with non-Hermitian system, there are cases that non-Hermitian Hamiltonian can be mapped to a Hermitian one via the transformation $H'=SHS^{-1}$, which is a case of pseudo-Hermiticity\cite{PhysRevLett.121.086803}.  Due to the non-Hermiticity of $H$, the mapping operator shall be non-unitary. However, the conventional mapping operator is time irrelevant, now for periodically driven system, we expanded it to be a time dependent $S(t)$, which induce the definition of dynamical pseudo-Hermiticity. The central part of non-Floquet theory features a temporal transformation of the Floquet states, which is closely related to the dynamical pseudo-Hermiticity.

To be specific, we consider to transform the Floquet states in a non-unitary way, $\Psi'(t)=S(t)\Psi(t)$, and the transformed Floquet states are assumed to be of a time periodic Hermitian system, with $U'(T)\Psi'(t)=e^{-i\epsilon' t}\Psi'(t),\epsilon' \in \mathrm{R}$, where $U'(t)=\mathcal{T} e^{-i\int^{T}_0 H'(t) dt}$  is the time evolving operator of the the Hermitian Hamiltonian. Combining the assumptions above with the Floquet theorem, the time periodic non-Hermitian Hamiltonian can be related to a Hermitian one in the following way:
\begin{eqnarray}
H'=SHS^{-1}+i\dot{S}S^{-1}
\end{eqnarray}
As the time periodic Hamiltonian $H'$ for the transformed Floquet states are assumed to be Hermitian, then dynamical pseudo-Hermiticity is supposed to be achieved (see Supplementary materials for details):
\begin{equation}
H^{\dagger}-\theta_1 H (\theta_1)^{-1}=i(\theta^{\dagger}_2+\theta_1 \theta_2\theta_1^{-1})
\end{equation}
where $\theta_1=S^{\dagger}S, \theta_2=S^{-1}\dot{S}$. For the special case that mapping operator $S$ is time independent, equation (3) is reduced to the conventional pseudo-Hermiticity form. Besides, the time evolving operators of the two models can be connected to each other accordingly, where $U'(T)=S(T)U(T)S^{-1}(0)$.  The deformation operator is applied in a way  that a time dependent complex phase is attached to different lattice sites,  we can denote it as $S(t)=e^{\Gamma (t)}$, where $\Gamma(t)$ is a diagonal matrix. It can be  exhibited that the quasienergy of the periodically driven non-Hermitian, Hermitian model can be tuned to each other via:
\begin{eqnarray}
\epsilon'=\epsilon+i\frac{1}{T}[\Gamma(T)-\Gamma(0)]
\end{eqnarray}
Equation (4) implies that  the Floquet edge modes of the periodically driven non-Hemritian model can be predicted  from its dynamical pseudo-Hermiticity counterpart. Examples include a periodically driven non-Hermitian one dimensional bipartite chain. (see Fig. \textbf{1a}).
\begin{figure}[tb]
\centering
\includegraphics[width=0.48\textwidth]{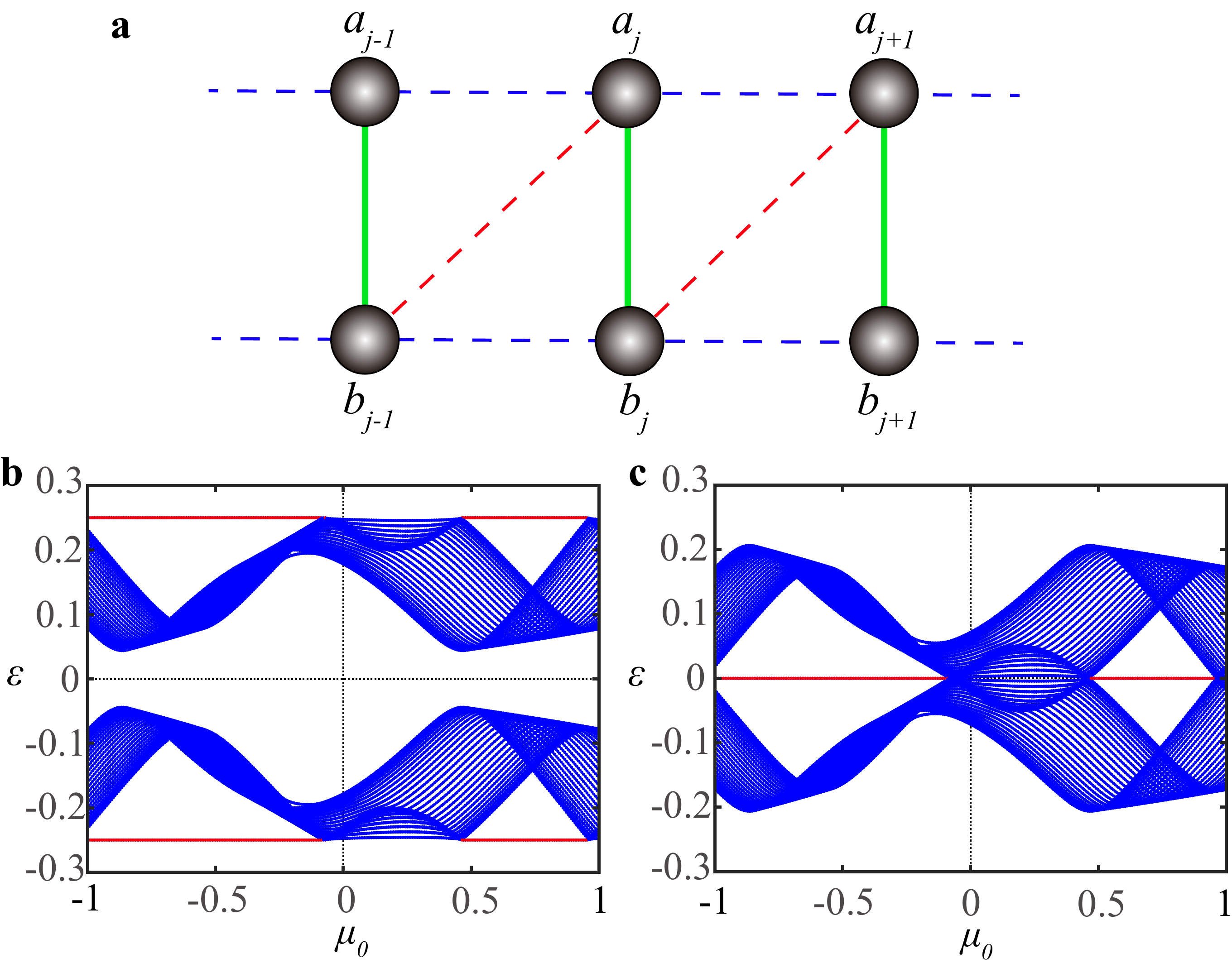}
\caption{\textbf{Time periodic non-Hermitian Hamiltonian and its dynamical pseudo-Hermiticity counterpart.} \textbf{a}, Schematic presentation of the time periodic non-Hermitian one-dimension bipartite chain.  Non-Hermiticity lies in both spacial and temporal asymmetric tunneling and also the dissipation indicated by the complex chemical potential. \textbf{b}, Quasienergy spectrum of the periodically driven non-Hermitian Hamiltonian in equation (5) with $\omega=0.5, r_1=0.05\omega, r_2=0.2\omega, v=-1.0\omega, q_1=-0.1\omega$ and $q_2=-0.4\omega$. \textbf{c}, Quasienergy spectrum of the periodically driven Hermitian model which is the dynamical  pseudo-Hermiticity counterpart of the model in equation (5) with $\omega =0.5,t_1=0.1\omega, t_2=1.0\omega$, and $p=-0.2\omega$ . Red lines depict the Floquet edge modes and blue lines correspond to the bulk states.} \label{fig1}
\end{figure}
\begin{figure}[htb]
\centering
\includegraphics[width=0.48\textwidth,height=0.57\textheight]{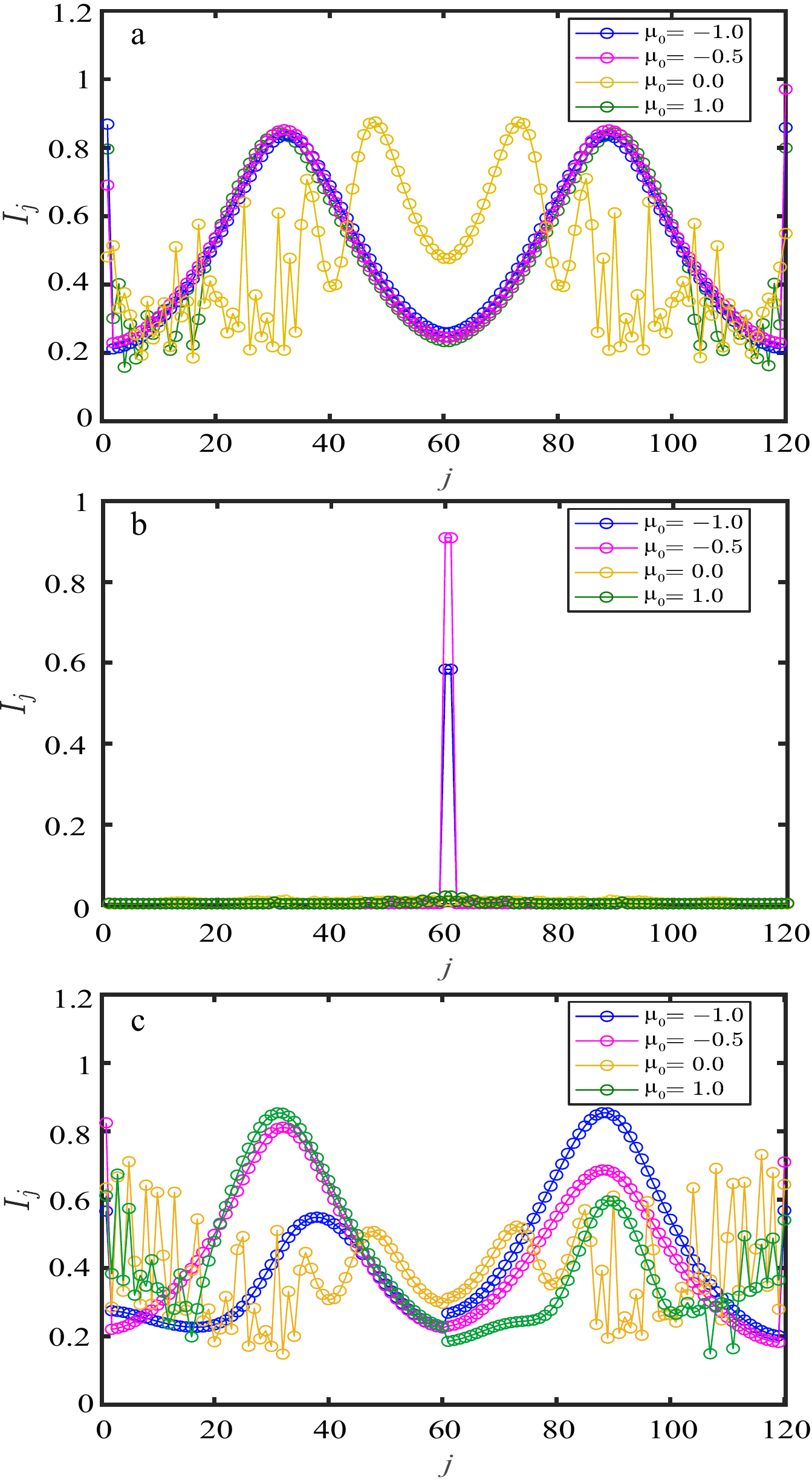}
\caption{\textbf{The localization factor $I_j$ for the $j th$  Floquet states of the periodically driven one dimensional bipartite chain.} \textbf{a}, The non zero localization factor for the model in equation (5) with driving period $[0,T]$, implies that apart from the Floquet edge modes but all bulk states are localized as well. \textbf{b}, The vanished localization factor denotes equally extended bulk states for the dynamical pseudo-Hermiticity counterpart of model (5). The peak for $\mu_0=-1.0, -0.5$ suggests the localized Floquet edge modes. \textbf{c}, For time periodic non-Hermitian chain, tuning the driving period to $[0.5\pi, T+0.5\pi]$, localization behavior differs drastically from that in Fig. \textbf{2a}. } \label{fig2}
\end{figure}
\begin{align}
H=&\sum_{j}f_1a_j^{\dagger}b_j+f_2b^{\dagger}_j a_j+g_1 b^{\dagger}_ja_{j+1}+g_2a^{\dagger}_{j+1}b_j\nonumber\\
&+p_1a^{\dagger}_ja_{j+1}+p_2a^{\dagger}_{j+1}a_j-p_1 b^{\dagger}_jb_{j+1}-p_2b^{\dagger}_{j+1}b_j\nonumber\\
&-\mu a^{\dagger}_ja_j+\mu b^{\dagger}_j b_j
\end{align}
where $a_j,b_j$ is the annihilation operator of sublattices on $jth$ site, and $L$ denotes the length of the chain. The intra-cell and inter-cell tunneling vary with time in a spatial-temporal asymmetric way that: $f_1\!=\!-r_1 e^{-i\omega t+2\cos{\omega t}}, f_2\!=\!-r_2 e^{i\omega t-2\cos{\omega t}}, g_1\!=\!v e^{i\omega t-2\cos{\omega t}}, g_2\!=\!v e^{-i\omega t+2\cos{\omega t}}, p_1=q_1\sin{\omega t}, p_2=q_2\sin{\omega t}$.  The system is dissipative with the complex chemical potential $\mu=(\mu_0+i\omega)\sin{\omega t}-\frac{\omega}{2}$.  Via the spatial-temporal non-unitary transformation: $S_j(t)=\bigoplus^{j=L}_{j=1} \left( \begin{smallmatrix}
             e^{\beta(j-1)-i\frac{\omega t}{2}+\cos{\omega t}} & 0 \\ 0 & e^{\beta j+i\frac{\omega t}{2}-\cos{\omega t}}
          \end{smallmatrix} \right)$, the dynamical pseudo-Hermiticity counterpart of the model in equation (5) can be obtained, whose Hamiltonian in momentum space reads $H(t)=\sum_{i=x,y,z} d_i(k)\cdot \sigma_i$,
\begin{figure*}[htb]
\centering
\includegraphics[width=0.97\textwidth]{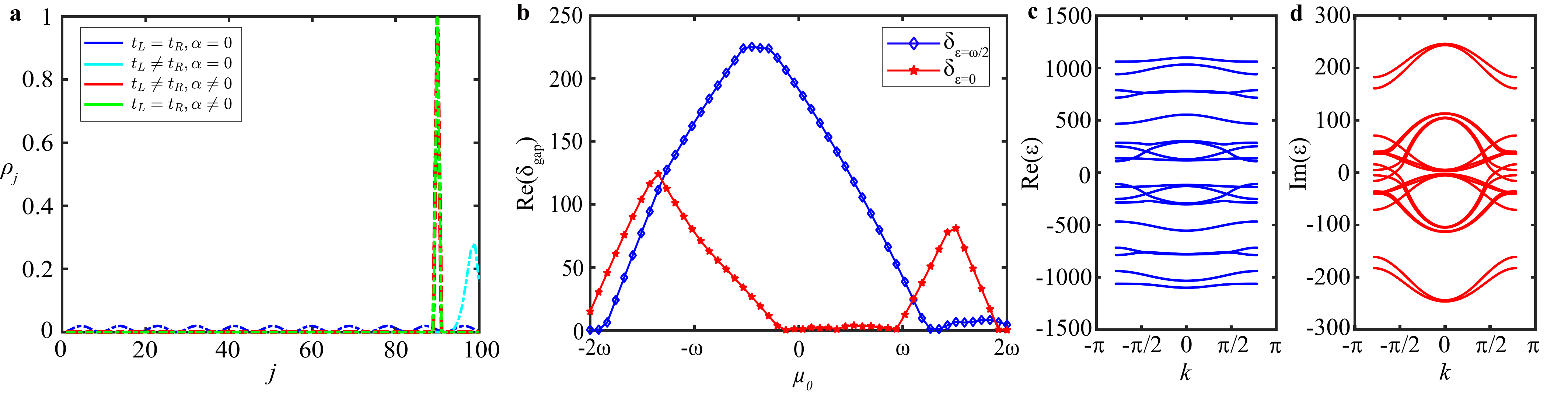}
\caption{ \textbf{The localization behavior of tight binding states and frequency space Floquet band spectrum.} \textbf{a}, With vanished external field, the tight binding states form plane waves (blue lines) for Hermitian model with symmetric tunneling, and form localized states piled up at one side (cyan lines) for the non-Hermitian model with asymmetric tunneling. With large enough external field, the tight binding states are of Wannier-Stark localization in either Hermitian or non-Hermitian cases and the eigenstate corresponding to the $90th$ eigenenergy is just localized at $90th$ lattice site. \textbf{b}, The Floquet $\frac{\pi}{T}$ gap amplitude (blue lines) of the time periodic non-Hermitian model described in  equation (5) and Floquet zero gap amplitude  (red lines) of the temporally deformed non-Floquet Hamiltonian in equation (7).  \textbf{c-d}, The Floquet bands of time periodic non-Hemitian Floquet Hamiltonian in equation (5). Apart from the real spectrum shown in Fig. \textbf{3c}, there also exists complex part in Fig. \textbf{3d} with high frequency driven field, $\omega=200$.}
\end{figure*}
where $d_z(k)\!\!=\!\!(2p\cos{k}-\mu_0)\sin{\omega t}, d_y\!=\!-t_2\sin{k}, d_x(k)\!=\!-t_1-t_2\cos{k}$ and the $\mu_0, p, t_1(t_2)$ correspond to the staggered chemical potential, the inter-cell tunneling between the same sublattice, the intra (inter)-cell tunneling between different sublattice respectively. It can be inferred from the form the transformation operator that the Floquet edge modes for time periodic Hermitian and non Hermitian models are related to each other with a quasienergy shift $\delta \epsilon=\frac{\omega}{2}$, see also Figs. \textbf{1b-1c}.  In addition, the Floquet edge modes in the Hermitian model are protected by chiral symmetry via $\mathcal {C} U\mathcal{C}^{-1}=U^{-1}$ with $\mathcal{C}=\delta_{j,j'}(\sigma_z)_{r,r'}$, where $r,r'$ is the sublattice index (see supplementary materials for details). The original periodically driven non-Hermitian model can also be chiral symmetry protected in the same way as long as $[\mathcal{C}, S_j(t)]=0, t\in [0,T]$, which is assured with our model in equation (5). \\
\textbf{Generalized quasienergy and starting point relevant localization.}
In the non-Bloch band theory, along with the spatial deformation of eigenstates of non-Hermitian tight binding model, the Brillouin Zone is also altered to lie in the complex plane. In the same spirit, together with the temporal non-unitary deformation of Floquet states, the Floquet quasienergy shall also be manipulated, which can be implied by the cases shown in equation (4). Here, we rename the quasienergy corresponding to the transformed Floquet states as generalized quasienergy. Typically, the quasienergy regarding to the original Floquet states can either be real or complex, which is closely related to the details of driving scenario (see supplementary materials for details).

The generalized quasienergy benefits to restore discrete time translational symmetry for the Floquet states via $\Psi'(NT)=e^{-i\epsilon' NT}\Psi'(0)$. It shall be noted that the generalized quasienergy  differs from  the original quasienergy only by an energy shift, and can not generate other complex band gaps. Hence, the topological features defined via  the winding of quasienergy and generalized quasienergy are the same. However, to utilize generalized quasienergy and to restore the discrete time translational symmetry do matters in other non-topological aspects. Before making an illustration, we notice that the spatially-temporally deformed Floquet states may have total different localization behavior from the original Floquet states, which can numerically described by the localization factor $I_j$ of the real space Floquet states $\Psi_j$, where $I_j=\sum^{2L}_{m=1}|\Psi_j(m)|^4-\frac{1}{2L }(\sum^{2L}_{m=1}|\Psi_j(m)|^2)^2, j\in[0,2L]$. For equally extended Floquet states, we have $I_j\varpropto \frac{1}{2L}$ vanished in the thermodynamic limit.  In Figs. \textbf{2a-2b}, we see that for time periodic non-Hermitian model, all states are localized with nonvanished localization factor. A quite striking feature of time periodic non-Hermitian system lies in that, in sharp contrast to the Hermitian driving scenario, different starting points of the driving period can result to drastically different localization behaviors. To account this phenomenon, we notice that turning the driving period from $[0, T]$ to $[t_0, T+t_0]$ amounts to the transformation: $U(t_0+T, t_0)=U(t_0,0) U(T,0) U^{-1}(t_0,0)$. Since $U(t_0,0)$ is spatially non-unitary, the Floquet states of the two driving scenario, are therefore differently localized. The different localization above, of the same spirit, can also be achieved with tuning the constant phase at each instantaneous time of a driving period (see supplementary materials for details). In a practical view, above effect can be utilized reversely to detect quantum phases of oscillating field.  The initial time (constant phase) relevant localization can also be manifested in an experimental aspect in ways that the detection of photon occupation number in photonic system with unidirectional transport may not be stable but varies with constant phases. Here, we point it out that by substituting the Floquet operator with the transformed one of generalized quasienergy, effects above can be eliminated and stable (phase irrelevant) photon occupation number can be obtained.\\
\textbf{non-Floquet theory.}
Non-Floquet theory  centers the temporal non-unitary deformation on Floquet states, effects of which get manifested with the frequency space Floquet Hamiltonian, with the aim of restoring Wannier-Stark localization.  To begin with, we restate the Floquet theorem  as $\Psi(t)\!=\!e^{-i\epsilon t}\phi (t); \phi(t)\!=\!\phi(t+T)=\!\sum_m e^{im\omega t}\tilde{\phi}(m)$. The Floquet Schrodinger equation  in frequency space reads: $\sum_p \tilde{h}_p\tilde{\phi}(m-p)\!\!=\!\!(\epsilon-m\omega)\tilde{\phi}(m)$, where $\tilde{h}_p\!\!=\!\!\frac{1}{T}\int^{T}_0 H(t) e^{-ip\omega t} dt$. The temporal asymmetric non-Hermitian terms at each instantaneous time may result in  asymmetric tunneling terms in the frequency space lattice model via $\tilde{h}_{p}\neq \tilde{h}^{\ast}_{-p}$. To retain Hermiticity, $\tilde{\phi}(m)$ can be replaced with $\tilde{\phi'}(m)=e^{\Gamma} \tilde{\phi} (m)$. In the following, we shall focus on two cases. For case $\mathrm{I}$, where $\Gamma$ is irrelevant with time, say $\Gamma=\eta m$, above replacing  is not necessary and Floquet bands corresponding to $\tilde{\phi}(m),\tilde{\phi'}(m)$ can  both give an appropriate description on topological phases with high frequency driving; For cases $\mathrm{II}$, where $\Gamma$ is time relevant, say $\Gamma=\Gamma(\omega t)$,  transformation above can be necessary in predicting Floquet topological phase transitions via the generation of Floquet band gaps.

In detail, for cases $\mathrm{I:}$ $\Gamma=\eta m$, recovering Hermiticity in the tunneling part of Floquet Hamiltonian requires:
\begin{equation}
\tilde{h}_{p} e^{-\eta p}= (\tilde{h}_{-p} e^{\eta p})^{\ast}
\end{equation}
Equation (6) is achievable with harmonically driven systems in high frequency limit, where $\omega \gg \Lambda$, $\Lambda$ is the band energy scale.  Remarkably, we exhibit that despite the asymmetric tunneling in frequency space, quasienergy $\epsilon$ can still be real and forms Wannier-Stark ladders $\frac{\epsilon}{\omega}-m=0$, as long as the complex band energy scale of the non-Hermitian Floquet Hamiltonian ($\tilde{h}_p$) is incomparable to the amplitude of the driving frequency. In addition, similar to the Wannier-Stark localization in time periodic Hermitian systems, eigenstate manifolds corresponding to the $ith$ eigenenergy manifolds of non-Hermitian Floquet Hamiltonian, perform localization on $ith$ frequency lattice site. To illustrate demonstration above, we consider a one dimensional chain with asymmetric tunneling and also a uniform external field, whose Hamiltonian reads: $H\!\!=\!\!\sum^{l=N-1}_{l=1} (t_L c^{\dagger}_lc_{l+1}+ t_R c^{\dagger}_{l+1}c_{l})+\sum^{l=N}_{l=1}\alpha l c^{\dagger}_l c_l$. $\alpha$ denotes the strength of the external field, and $N$ is the length of the chain. When we regard the asymmetry in tunneling as an effect of complex flux, model above can be mapped to a Hermitian one: $H'\!\!=\!\!\sum^{l=N-1}_{l=1} (\sqrt{t_L t_R} d^{\dagger}_ld_{l+1}+ h.c.)+\sum^{l=N}_{l=1}\alpha l d^{\dagger}_l d_l$. Analytically, via the recursion equation, it can be obtained $c_l=e^{-\beta l}(A Y_{\frac{\epsilon}{\alpha}-l}(\frac{2\sqrt{t_Lt_R}}{\alpha})+B J_{\frac{\epsilon}{\alpha}-l}(\frac{2\sqrt{t_Lt_R}}{\alpha}))$ for the original non-Hermitian chain, where $J_{\nu}(x), Y_{\nu}(x)$ are Bessel functions of first and second kind, and $A,B $ are constants determined by the boundary conditions, $e^{\beta}=\sqrt{\frac{t_L}{t_R}}$. Hence, it can be implied that in high field limit, $\alpha\gg 2\sqrt{t_Lt_R},\sqrt{\frac{t_R}{t_L}}$, non-vanished eigenstates on $lth$ eigenenergy ladder must be localized at $lth$ site. Thus, Wannier-Stark localization can win over the skin effects in high field limits, and in low field, conclusion reverses. Numerical results in Fig. 3\textbf{a} present the population rate of $jth$ eigenstate of the non-Hermitian chain  by $\rho_j=\mathrm{diag(\frac{|\phi_j\rangle\langle\phi_j|}{\langle\phi_j|\phi_j\rangle})}$, which  shows a sound agreement with the analysis above. In consequence, truncations of Hermitian and non-Hermitian Floquet Hamiltonian in high frequency limit are both appropriate to predict Floquet topological phases.

In case $\mathrm{II}:$ $ \Gamma=\Gamma(t)$. Wannier-Stark ladder and localization of eigenstates in frequency space can be broken for non-Hermitian Floquet Hamiltonian with a moderate complex band energy scale, $\lim_{\omega\rightarrow\infty} \frac{1}{\omega}|\tilde{h}_p|\neq 0$.  The numerical origin for break above lies in that Bessel functions of moderate variable are not maximally localized with zeroth order.  To cope this scenario, a temporal non-unitary deformation, $\Phi'(t)=e^{\Gamma (t)}\Phi (t)=e^{-i\epsilon' t}\phi'(t)$, has to be applied. We present that Wannier-Stark localization can be restored with the deformed Floquet states via (see supplementary materials for details):
\begin{eqnarray}
\begin{aligned}
(e^{\Gamma} H e^{-\Gamma} -i\partial_t)\phi'(t)&=(\epsilon'-i\dot{\Gamma}) \phi'(t)\\
\phi'(t)&=\sum_m e^{i m\omega t }\tilde{\phi}' (m)
\end{aligned}
\end{eqnarray}
where $H$ describes the time periodic non-Hermitian Hamiltonian, and $\epsilon' \in \mathrm{R}$. The term $i\dot{\Gamma}$ serve in two ways to restore Wannier-Stark localization in frequency space: 1) eliminating the complex chemical potential in frequency space. 2) reducing the Floquet bands energy scale and $|m-m'|\omega \gg|\Lambda|$ can be readily achieved.

As an  illustration of case $\mathrm{II}$, we stick to the time periodic  one dimensional bipartite chain presented in Fig. \textbf{1a}. Nevertheless, we start from the dynamical pseudo-Hermiticity counterpart and perform the non-unitary deformation reversely. For the sake of simplification, we remove the spacial deformation, and only consider a  temporal deformation:   $S_j(t)=\bigoplus^{j=L}_{j=1} \left( \begin{smallmatrix}
             e^{-i\frac{\omega t}{2}+\cos{\omega t}} & 0 \\ 0 & e^{i\frac{\omega t}{2}-\cos{\omega t}}
          \end{smallmatrix} \right) $.
The newly transformed periodically driven non-Hermitian model remains the same in a schematic picture, but the asymmetric intra cell tunneling turns to $f_1=-t_1 e^{-i\omega t+2\cos{\omega t}}, f_2=-t_1 e^{i\omega t-2\cos{\omega t}}$, inter cell tunneling between same sublattices remains the same form as the original Hermitian one, $p_1=p_2=p\sin{\omega t}$, inter cell tunneling between different sublattices and chemical potential are of the same form as time periodic non-Hermitian chain in equation (5). It can be inferred from equation (4) that the quasienergy of the newly deformed Floquet operator is the same as that of model in equation (5), and thus should be totally real. In Fig. \textbf{3c-3d}, the Floquet bands of the newly deformed periodically driven non-Hermitian chain is presented, and there exists complex part.  The inconsistence of the frequency space Floquet bands and the real space Floquet edge modes is due to the breaking of Wannier-Stark localization,  where the off-diagonal part of Floquet Hamiltonian is complex, and also comparable to the amplitude of driving frequency (see supplementary materials for details). Non-Floquet engineering can fix the inconsistence above by transforming reversely to the dynamical pseudo-Hermiticity part, where the tunneling part of the Floquet Hamiltonian changes to $(-if\cos{k}+i\frac{\mu_0}{2})\sigma_z$, Wannier-Stark localization can be retained with $\omega \gg |t_1+t_2|, f$ (see supplementary materials for details). In Fig. \textbf{3b}, the Floquet band spectrum obtained via non-Floquet engineering is real, and the generation of Floquet zero gap  is consistent with the emergence of real space Floquet edge modes. To sum up, the central part of non-Floquet engineering is to deform the Floquet states temporally to restore Wannier-Stark states in high frequency limit, which can be utilized to identify different Floquet topological phases.\\
\textbf{Discussion.}
We are confident that the results in this paper will have a strong impact on designing new sets of experimental realization of Floquet topological materials and fundamental non-Hermitian dynamical physics. On the  practical side, our theory can be utilized to analyse the Floquet edge modes in dissipative non-Hermitian photonic topological insulators or  superconductors with time periodic oscillating electric-magnetic fields. Besides, our theory can aid to construct detectors for quantum phases in time periodic non-Hermitian systems, as different constant phases at each instantaneous time can result to different localization behaviours. In addition, topological materials with non-Floquet engineering can also be achieved in cold atom systems by periodically shaking dissipative optical lattice. We suggest future experimental and theoretical study  to  carry out a systematic study on dimensional effect on topological properties in dissipative dynamical system, which can be achieved by shaking dissipative optical lattice of higher dimensions or in a time quasi-periodical way. We believe that our works set up a complete theory for analyzing non-Hermitian dynamical topological phases, leading to possible new generation of dissipative Floquet topological materials.\\
\textbf{Acknowledgement} \\
We are grateful to Zhong Wang, V. Dwivedi and Peng Xue for helpful discussion. This work was supported by the National Key R and D Program of China under grants No. 2016YFA0301500, NSFC under grants Nos.  61835013.

\noindent\bibliography{huanyu-2}

\end{document}


\renewcommand\figurename{Supplementary Figure}
\title{Supplementary Material for: Non-Floquet engineering in periodically driven non-Hermitian systems}
\author{Huan-Yu Wang $^1$, Xiao-Ming Zhao $^1$, Lin Zhuang $^2$ and  Wu-Ming Liu $^1$ }
\affiliation{$^1$ Beijing National Laboratory for Condensed Matter Physics, Institute of Physics, Chinese Academy of Sciences, Beijing 100190, China}
\affiliation{$^2$ School of Physics, Sun Yat-Sen University, Guangzhou 510275, People's Republic of China  }
\maketitle
\section{Dynamical Pseudo-Hermiticity}
Non-Floquet theory centers a temporal non-unitary transformation on Floquet states, and effects of which can be manifested via an example of dynamical pseudo-Hermiticity process. For the temporally transformed Floquet states, $\Psi'(t)=S(t)\Psi(t)$, with the original states satisfying $U(t)\Psi(t)=e^{-i\epsilon t}\Psi(t),\Psi(t)=e^{-i\epsilon t}\phi(t)$; The $\phi(t)$ is governed by the Floquet-Schrodinger equation
\begin{eqnarray}
(H(t)-i\partial_t)\phi(t)=\epsilon \phi(t)
\end{eqnarray}
By substituting the temporal transformation, we can obtain how the transformed  states evolve in time:
\begin{eqnarray}
(SHS^{-1}+i\dot{S}S^{-1})\Psi'(t)=i\partial_t\Psi'(t)
\end{eqnarray}
By assuming the transformed model, $\tilde{h}_{dph}=SHS^{-1}+i\dot{S}S^{-1}$ to be of time periodic Hermitian scenario, which can induce the dynamical pseudo-Hermiticity process, in detail:
\begin{eqnarray}
\begin{aligned}
\tilde{h}_{dph}&=\tilde{h}^{\dagger}_{dph}\\
(SHS^{-1}+i\dot{S}S^{-1})^{\dagger}&=(SHS^{-1}+i\dot{S}S^{-1})\\
(S^{-1})^{\dagger} H^{\dagger} S^{\dagger}-SHS^{-1}&=i(S^{-1})^{\dagger}(\dot{S})^{\dagger}+i\dot{S}S^{-1}\\
(S^{-1})^{\dagger} H^{\dagger} S^{\dagger}S-SH&=i (S^{-1})^{\dagger}(\dot{S})^{\dagger}S+i\dot{S}\\
S^{-1}(S^{-1})^{\dagger} H^{\dagger}S^{\dagger}S-H &=i S^{-1}(S^{-1})^{\dagger}(\dot{S})^{\dagger}S+iS^{-1}\dot{S}
\end{aligned}
\end{eqnarray}
Denote $\theta_1=S^{\dagger}S$, $\theta_2=S^{-1}\dot{S}$, then subequation (3) can be presented as:
\begin{eqnarray}
\begin{aligned}
(\theta_1)^{-1} (H)^{\dagger}\theta_1-H&=i {\theta_1}^{-1}(\dot{S})^{\dagger}S+i\theta_2\\
&=i {\theta_1}^{-1}(\dot{S})^{\dagger}(S^{-1})^{\dagger}S^{\dagger }S+i\theta_2\\
&=i(\theta_1)^{-1}\theta_2^{\dagger}\theta_1+i\theta_2\\
H^{\dagger}-\theta_1 H (\theta_1)^{-1}&=i(\theta_2^{\dagger}+\theta_1 \theta_2 (\theta_1)^{-1})
\end{aligned}
\end{eqnarray}
For the static case, where $\theta_2=0$ and $H^{\dagger}=\theta_1 H (\theta_1)^{-1}$ just present the pseudo-Hermiticity form. As a expansion for time-dependent transformation, we rename $H^{\dagger}-\theta_1 H (\theta_1)^{-1}=i(\theta_2^{\dagger}+\theta_1 \theta_2 (\theta_1)^{-1})$ as dynamical pseudo-Hermiticity.
\section{Chiral symmetry for periodically driven one dimensional bipartite chain}
The periodically driven system can possess Floquet $0,\pm\frac{\pi}{T}$ edge modes with chiral symmetric time independent effective Hamiltonian, which is equivalent to $\Gamma U \Gamma=U^{-1}$, $\Gamma$ is the chiral symmetric operator. It is found that chiral symmetry can be identified when $U(T,0)=U(T,t_1)U(t_1,0)$, $\Gamma U^{\dagger}(T,t_1)\Gamma=U(t_1,0)$ \cite{JK}. For harmonically driven one dimensional bipartite chain $H'$, which is also the dynamical pseudo-Hermiticity counterpart of the model presented by equation (5) of main article, the Hamiltonian at each instantaneous time  in momentum space can be expressed as: $H'(k)=\sum_{i=x,y,z}d_i(k,t)\cdot\sigma_i$.  When $d_x(k,t),d_y(k,t)$ are even functions of time t, zero constant chemical potential ( or the chemical potential as an odd function of time t) can result in chiral symmetric Floquet operator with $\Gamma=\sigma_z$. Indeed both two cases above can lead to $\Gamma H(-t)\Gamma=-H(t)$, which is sufficient to preserve chiral symmetry of the evolving operator \cite{Balabanov}:
\begin{eqnarray}
\begin{split}
U(0,\frac{T}{2})=\Gamma U(0,-\frac{T}{2})\Gamma=\Gamma U^{\dagger}(\frac{T}{2},T) \Gamma
\end{split}
\end{eqnarray}
Thus, we can denote the Floquet operator as  $U_1=\Gamma U^{\dagger}(\frac{T}{2},T) \Gamma U(\frac{T}{2},T)=e^{-i H^1_{eff} T}$ or $U_2=U(\frac{T}{2},T)\Gamma U^{\dagger}(\frac{T}{2},T) \Gamma=e^{-iH^2_{eff} T}$. Considering the chiral symmetry,  we can express:
\begin{eqnarray}
H^{1(2)}_{eff}=\left( \begin{smallmatrix} 0 & q_{1(2)} (k) \\ q^{-1}_{1(2)}(k)& 0\end{smallmatrix} \right)
\end{eqnarray}
and the corresponding winding number is:
\begin{eqnarray}
\begin{split}
W_{1(2)}=\frac{i}{2\pi}\int^{2\pi}_0 q^{-1}_{1(2)}(k) dq_{1(2)}(k)
\end{split}
\end{eqnarray}
\begin{figure*}
\includegraphics[width= 0.7\textwidth,height=0.27\textheight]{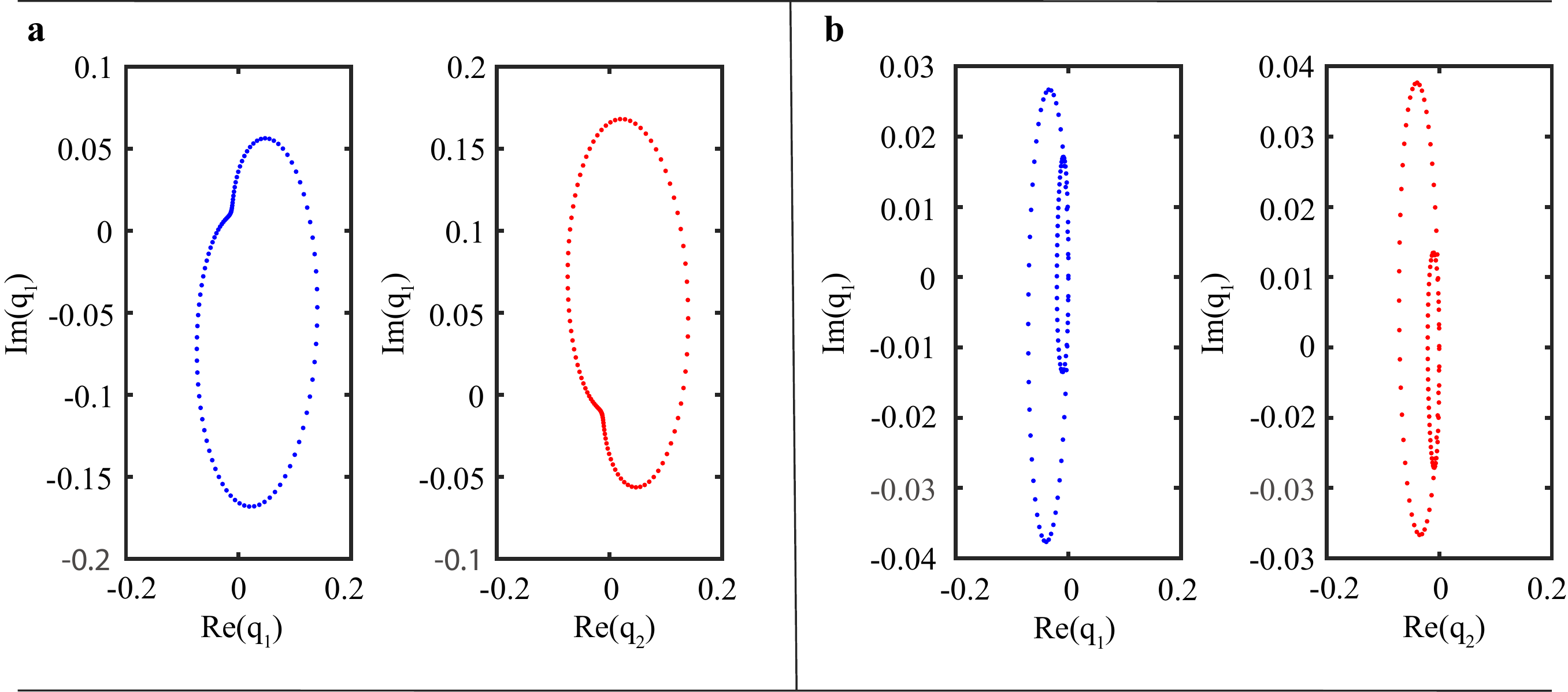}
\caption{(Color online). The $\{\mathrm{Re(q_{1(2)})},\mathrm{Im(q_{1(2)})}\}$ may circle around as the Bloch wave vector goes through the Brillouin zone. \textbf{a}, The one circle with no back stepping process at $\mu_0=-1$ indicate $W_{1(2)}=1$ and the presence of Floquet edge modes. \textbf{b}, The double circles with  back stepping process at $\mu_0=0$ indicates $W_{1(2)}=0$ and the absence of Floquet edge modes.  Results above consist with Floquet edge modes in Fig. \textbf{1c}.} \label{fig1}
\end{figure*}
\noindent The existence of the Floquet $0,\frac{\pi}{T}$ edge modes can be inferred by:
\begin{eqnarray}
\nu_0=\frac{W_1+W_2}{2},\quad\nu_{\pm\frac{\pi}{T}}=\frac{W_1-W_2}{2}
\end{eqnarray}
In our cases, Supplementary Figure 1 suggests that at $\mu_0=-1,0$, $W_1$ is the same as $W_2$. Hence, there shall be no Floquet $\pm\frac{\pi}{T}$ end modes, and $\nu_0$ is the same as $W_{1(2)}$ which implies  the existence of  Floquet zero end modes at $\mu_0=-1$.  Results above are in consistence with Floquet edge modes of the main article, which denote the bulk boundary correspondence.

\section{Floquet quasienergy for time periodic non-Hermitian systems}
Although the time periodic non-Hermitian Hamiltonian at each instantaneous time may hold complex energy, however the total Floquet quasienergy, which is closely related to the driven scenario, can be real. To illustrate how non-Hermiticity may influence the Floquet quasienergy, we concentrate on a seven  step quench process (see the schematic presentation in  Supplementary Figure \textbf{2a}), of which the Hamiltonian can be described as:
\begin{eqnarray}
H(k,t)=-\sum^{n=7}_{n=1}(J^1_n(t)e^{i\mathbf{b_n\cdot k}}\sigma^{+}+J^2_n(t)e^{-i\mathbf{b_n\cdot k}}\sigma^{-})+\Gamma\sigma_z
\end{eqnarray}
When the driven scenario is a Hermitian one with symmetric tunneling at each step, $J^1_n=(J^2_n)^{\ast}=J, n=1,2\cdots 7$, Floquet bands are fully degenerate with purely real quasienergy $\epsilon=\frac{\pi}{T}$ (see Supplementary Figure \textbf{2b-2c}). Next, we turn on the non-Hermitian process by setting $J^1_1=J+r, J^{2}_1=J-r$, and from Supplementary Figure \textbf{2d}, it can be seen that the imaginary part of Floquet quasienergy spectrum splits into two flat bands. Numerically, we have shown that the Floquet quasienergy can be manipulated back to be purely real, when the same asymmetric tunneling process is applied at quench step four. To get a more illustrative picture, staggered gain-loss process  is considered in Supplementary Figure \textbf{2e-2f} ($\mathrm{Im(\Gamma)}\neq 0$), and results show that the gap closing points can be changed with various gain-loss amplitude, leading to a richer topological phase diagram. In Supplementary Figure \textbf{2g}, numerical results show that, apart from manipulating the amplitude of gain-loss, tuning the asymmetric tunneling at different quench processes can also give birth to diverse gap closing points.  Analytically, we can judge the Floquet quasienergy to be real for a parity-time  ($\mathcal{PT}$) symmetric time independent effective hamiltonian and
\begin{eqnarray}
\mathcal{PT} U(k,T) \mathcal{PT}^{-1}=U^{-1}(k,T)
\end{eqnarray}
where $U(k,T)$ is the time evolving operator. If  $U(k,T)\Psi=e^{-i\epsilon T}\Psi, \mathcal{PT} \Psi=C\Psi$, then the quasienergy shall be real. It needs to be point out that time independent effective Hamiltonian can not inherit parity symmetry from the Hamiltonian at each instantaneous time, which is related to the details of driven scenario.
\begin{figure*}
\includegraphics[width= 0.7\textwidth]{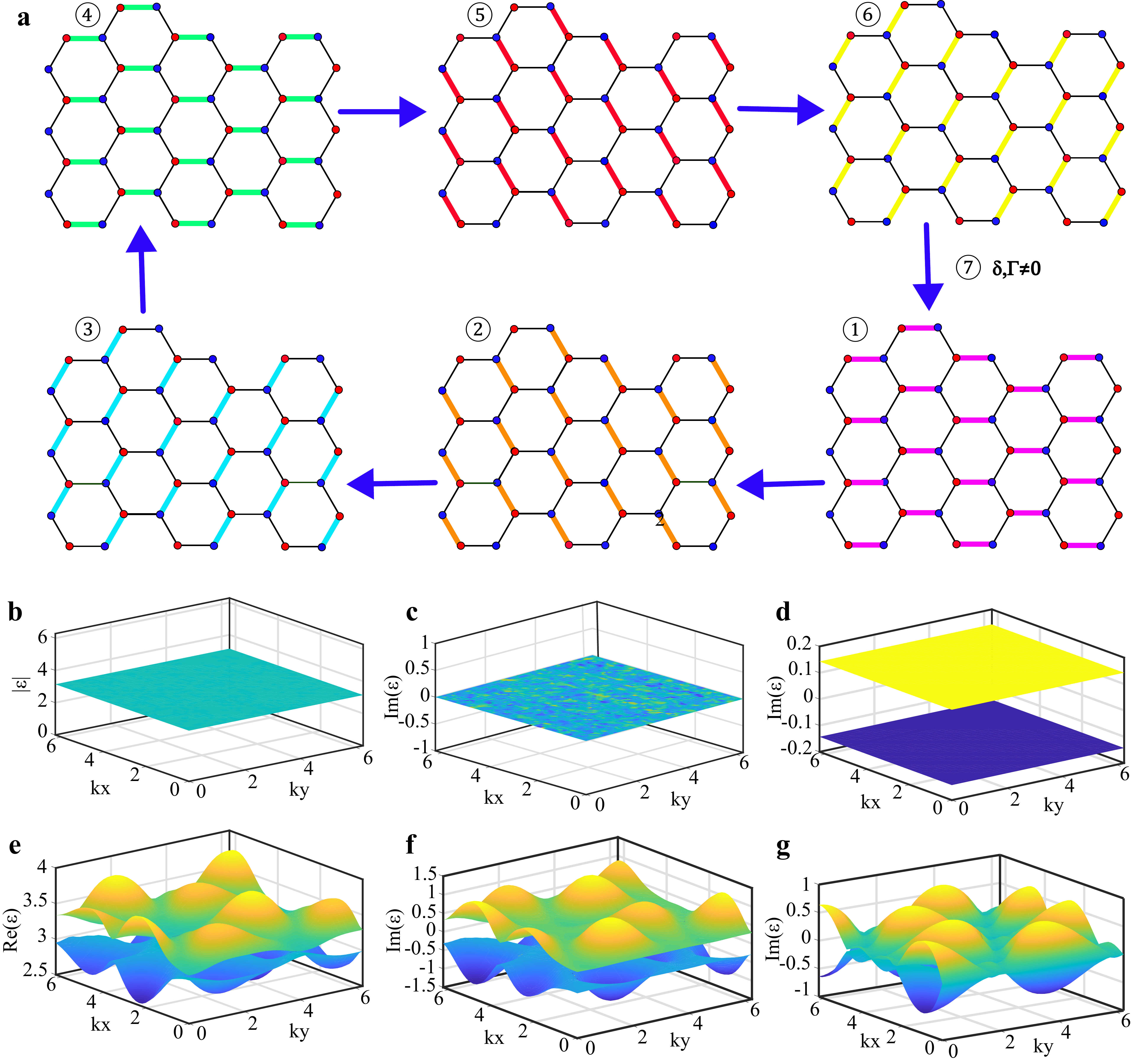}
\caption{(Color online). \textbf{a,} A schematic picture for the seven step quench process, and tunnelings are only allowed on colored bonds with the periodicity  set to be $T=1$. \textbf{b-c,} Fully degenerate real Floquet spectrum is present with symmetric tunneling $J^{1}_n=(J^{2}_n)^{\ast}=J=\frac{7\pi}{2}, \Gamma=0$. \textbf{d,} When the asymmetric tunneling is added $J^1_1=J+r, J^2_1=J-r, r=0.5\pi$, the Floquet spectrum  remains the same in the real part and splits into two flat bands for the imaginary part. When the same asymmetric tunneling is also exerted on step four, the imaginary part of Floquet quasienergy disappears. \textbf{e-f,} The staggered gain-loss is turned on with $\Gamma=1+2i$. \textbf{g,} Floquet quasienergy spectrum with diverse gap closing points can also be obtained by tuning the asymmetric tunneling in different quench processes, $J^1_{1,2}=J+r_1, J^2_{1,2}=J-r_1, J^{1}_{3,4}=J-r_1, J^2_{3,4}=J+r_1$. } \label{fig1}
\end{figure*}

\begin{figure}
\includegraphics[width= 0.9\textwidth,height=0.38\textheight]{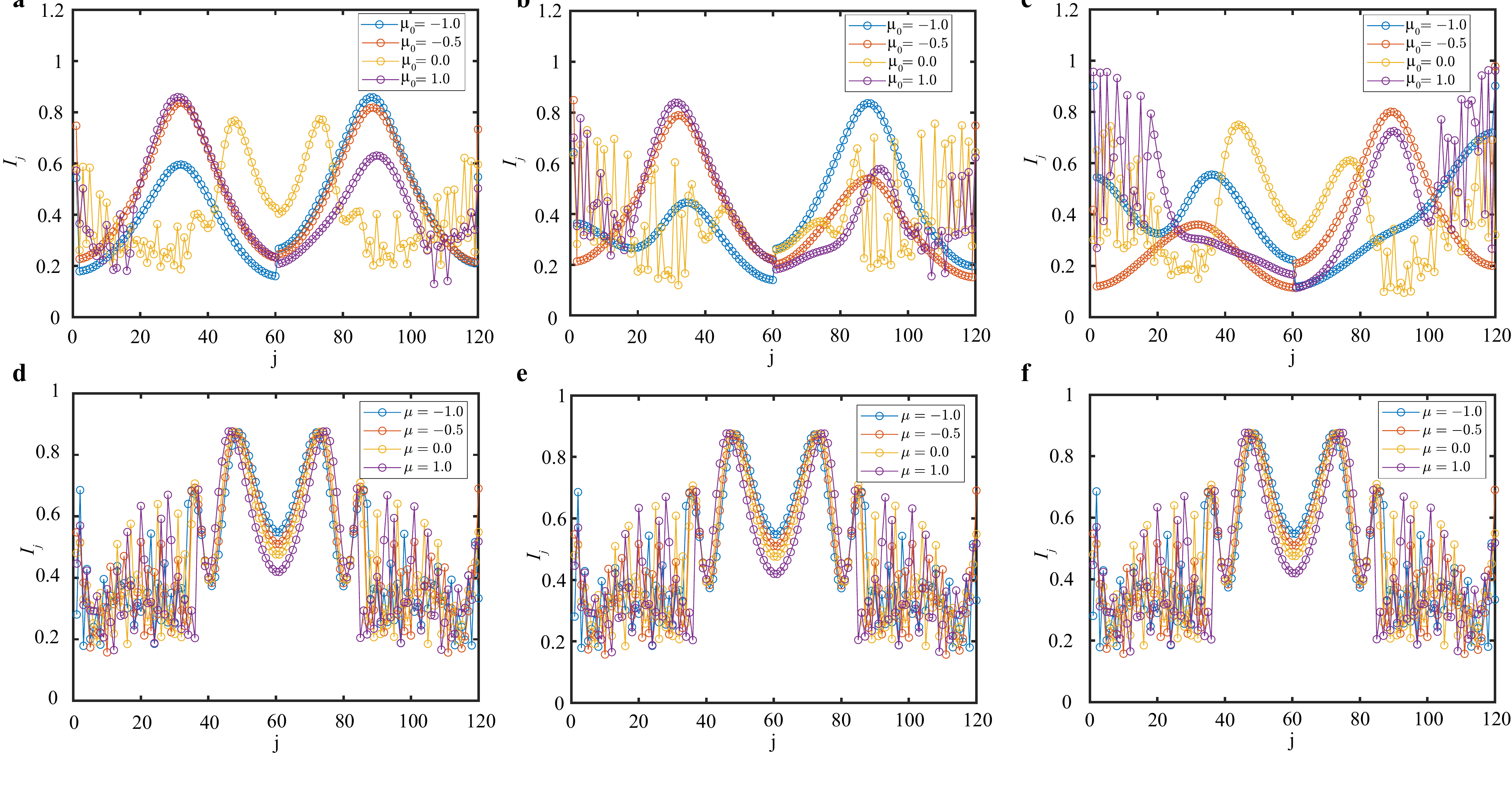}
\caption{(Color online). \textbf{a-c,} The localization factor of the Floquet states for periodically driven non-Hermitian one dimension bipartite chain with different starting points (constant phase) and $\omega=0.5$. \textbf{a,} $\phi=0.3\pi$, \textbf{b,} $\phi=0.6\pi$, \textbf{c,} $\phi=1.2\pi$.  \textbf {d-f,} The localization factor of the time periodic non-Hermitian Floquet states with different starting time (constant phase) and $\omega=20$. \textbf{d,} $\phi=0.3\pi$, \textbf{e,} $\phi=0.6\pi$, \textbf{f,} $\phi=1.2\pi$ } \label{fig1}
\end{figure}

\section{different localization induced by starting points of driving period}
In periodically driven Hermitian system, constant phase at each instantaneous time is usually ignored as it contributes to no difference in Floquet quasienergy spectrum but just a unitary shift in Floquet states, which can be gauged out by choosing starting time of driving period properly. In detail, for a time periodic Hermitian Hamiltonian: $H(t)=H_0+H_1(t+\phi)$. $H_0$ is the static part, and $\phi$ is a constant phase.  The Floquet operator is $U(T,0)=\mathcal{T} e^{\int^T_0 H(t) dt}$. By changing integral variable $t'=t+\phi$, we can express the Floquet operator as $U(T+\phi,\phi)$, which is the same as changing driving initial time. By the way, we have:
 \begin{equation}
U(T+\phi,\phi)=U(T+\phi,T)U(T,0)U(0,\phi)=U(\phi,0)U(T,0)U^{-1}(\phi,0)
\end{equation}
Results above suggest that constant phase at each instantaneous time only result in a transformation in Floquet operator with transforming operator $U(\phi,0)$, which can usually be ignored in the Hermitian scenario. However, with periodically driving non-Hermitian system, this is a non-unitary transformation. Although the Floquet quasienergy spectrum remains the same, the localization behavior of the Floquet states change drastically, skin effect included. In Supplementary Figure 3, we obtain the localization factor with different constant phase $\phi=0.3\pi,0.6\pi, 1.2\pi$, which differs drastically. As we turn to high frequency driving cases, it can be seen that the different localization behavior induced by constant phase is suppressed, which is due to that in fast driving scenario, the initial phases equal to an integer amount of  driving period, and for off resonance cases, the non-unitary shifts induced by $U(\phi,0), U^{-1}(\phi,0)$ get eliminated.  Meanwhile results above also implicate that in contrast to the Hermitian driving scenario, the choice of initial time plays an important role in deciding the localization behavior.
\section{Wannier-Stark localization of deformed Floquet states}
Before going further into the Wannier-Stark localization in frequency space, we first explore what the Wannier state regarding to the non-Bloch band theory looks like, to start with, we have:
\begin{eqnarray}
\begin{aligned}
W(x-a)&=\sum_k e^{-ik\cdot a}\Psi(x)\\
&=(det(T))^{\frac{x}{2}}\sum_k e^{ik(x-a)}\sum_m e^{im\frac{2\pi}{a} x}\tilde{\phi}(m)
\end{aligned}
\end{eqnarray}
\begin{figure}
\includegraphics[width=0.4\textwidth]{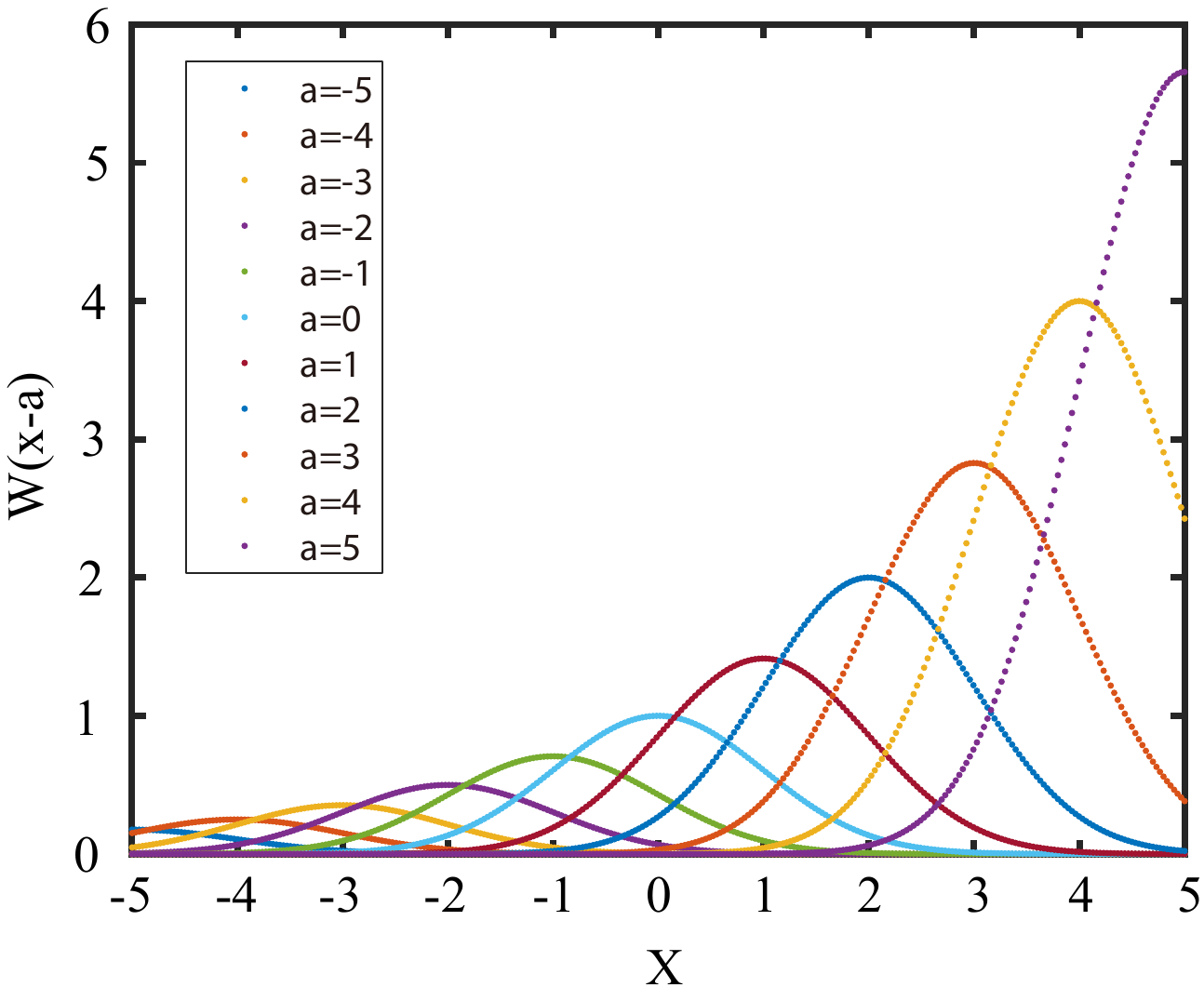}
\caption{(Color online). Representation of  Wannier states in non-Bloch band theory. } \label{fig1}
\end{figure}
where $T$ is the transfer matrix. Results above show that the Wannier states in non-Bloch band theory differ from the conventional Wannier states by a site dependent amplification. The physical origin for this lies in that, tight binding states in non-Bloch band theory do not equally localized for each lattice potential, instead whose localization behavior can be decided by the whole Hamiltonian and boundary conditions.  To get an illustrative picture, we use Gaussian function to simulate the Wannier state localized at $x=a$ in deep potential.  From Supplementary Figure 4, it can be seen that the skin effect wins over the Wannier localization and plays a dominant role. Meanwhile, conventionally for topological non-trivial one dimensional chain, as the Bloch wave vector goes around a close parameter space, there can exist $2n\pi$ phase accumulated in tight binding states, which corresponds to $n$ site moving of Wannier center. However, Wannier localization differs site by site in non-Hermitian tight binding model,  $2n\pi$ accumulated phase in tight binding states can not just account to a integer movement of Wannier center, which indeed suggests that apart from topological  pumping charge, there also exists charge exchanging process with the external environment.

For the frequency space Floquet Hamiltonian, the tight binding model differ from the Hermitian lattice model by a uniform external field, which in high frequency limit, can win over skin effect and take the leading place.  As an illustration of non-Floquet engineering, we do the time dependent non-unitary deformation on the Floquet states:
\begin{eqnarray}
\begin{aligned}
&\Phi'(t)=e^{\Gamma(t)}\Phi(t)=e^{-i\epsilon' t}\phi'(t)=e^{-i\epsilon t+\Gamma (t)}\phi(t)
\end{aligned}
\end{eqnarray}
Consider a complete driving period, we shall have the non-deformed and deformed Floquet operator related via:
\begin{eqnarray}
U'(T)=e^{\Gamma(t+T)}U(T) e^{-\Gamma(t)}
\end{eqnarray}
The $\phi(t)$ and $\phi'(t)$ obey different time evolving equation:
\begin{eqnarray}
\begin{aligned}
(H-i\partial_t)\phi(t)&=\epsilon\phi(t)\\
(e^{\Gamma(t)} H(t) e^{-\Gamma (t)}-i\partial_t) \phi'(t) &=(\epsilon'-i\dot{\Gamma})\phi'(t)
\end{aligned}
\end{eqnarray}
By expressing $\phi(t)=\sum_m e^{im\omega t}\tilde{\phi}(m),\phi'(t)=\sum_n e^{in\omega t}\tilde{\phi'}(n)$, we can have:
\begin{eqnarray}
\begin{aligned}
&\frac{1}{T}\int^{T}_0 \sum_m H(t) e^{i(m-m')\omega t} \tilde{\phi}(m)+m'\omega \tilde{\phi}(m')=\epsilon\tilde{\phi}(m')\\
&\frac{1}{T}\int^{T}_0 \sum_n e^{\Gamma(t)} H(t) e^{-\Gamma (t)} e^{i(n-n')\omega t} \tilde{\phi}(n)+n'\omega \tilde{\phi}(n')=(\epsilon'-i\dot{\Gamma})\tilde{\phi}(n')
\end{aligned}
\end{eqnarray}
Assuming  the deformation do not disturb the driving frequency $H'(t)=e^{\Gamma(t)} H(t) e^{-\Gamma (t)}, H'(t)=H'(t+T)$, By Fourier transformation, $H(t)=\sum_p \tilde{h}_p e^{ip\omega t}, H'(t)=\sum_q \tilde{h'}_q e^{iq\omega t}$, it can be achieved:
\begin{eqnarray}
\begin{aligned}
&\sum_m \tilde{h}_{m'-m}\tilde{\phi}(m)+m'\omega \tilde{\phi}(m')=\epsilon \tilde{\phi}(m')\\
&\sum_n \tilde{h'}_{n'-n}\tilde{\phi'}(n)+n'\omega \tilde{\phi}(n')=(\epsilon'-i\dot{\Gamma})\tilde{\phi}(n')
\end{aligned}
\end{eqnarray}
When the system is of non-Hermitian driven scenario, Wannier-Stark ladder of eigenenergy can not be applied in following cases:
\begin{eqnarray}
\tilde{h}_{m-m'}\neq (\tilde{h}_{m'-m})^{\ast}, \quad \lim_{\omega\rightarrow\infty}\frac{1}{\omega} |\tilde{h}_{m-m'}|\neq 0
\end{eqnarray}
However, by carefully choosing the form of  $\Gamma(t)$, the Wannier-Stark localization can be restored as long as:
\begin{eqnarray}
\frac{1}{\omega} (|\tilde{h'}_{n-n'}|+i\dot{\Gamma})\approx0
\end{eqnarray}
As an illustration, we review the example in case $\mathrm{II}$ of the main article, the time periodic non-Hermitian Hamiltonian reads by  $H=H_0+H_t$ and:
\begin{eqnarray}
H_{0}=\left( \begin{smallmatrix} \frac{\omega}{2} & 0 \\ 0 & -\frac{\omega}{2} \end{smallmatrix} \right)
\end{eqnarray}
\begin{eqnarray}
H_{t}=\left( \begin{smallmatrix} [2p\cos{k}-(\mu_0+i\omega)]\sin{\omega t} & (-t_1-t_2 e^{-ik})e^{-i\omega t +2\cos{\omega t}} \\ (-t_1-t_2 e^{ik})e^{i\omega t-2\cos{\omega t}} & [-2p\cos{k}+(\mu_0+i\omega)]\sin{\omega t} \end{smallmatrix} \right)
\end{eqnarray}
Generally, $H_t$ contributes to the off-diagonal term in Floquet Hamiltonian. By substituting subequation (21) into the first term of subequation (16), it can be obtained:
\begin{eqnarray}
\begin{aligned}
\hat{H}^{off-diagonal}&=\frac{1}{T}\int^{T}_0 \sum_m H_t e^{i(m-m')\omega t} dt\\
\hat{H}^{off-diagonal}_{1,1}&=-\hat{H}^{off-diagonal}_{3,3}=\frac{1}{T}\int^{T}_0 (2p\cos(k)-(\mu_0+i\omega))\sin{\omega t} e^{i(m-m')\omega t} dt\\
&=\sum_m (ip\cos{k}+\frac{w}{2}-\frac{i\mu_0}{2})[\delta_{m=m'+1}-\delta_{m=m'-1}]
\end{aligned}
\end{eqnarray}
\begin{eqnarray}
\begin{aligned}
\hat{H}^{off-diagonal}_{1,2}&=\frac{1}{T}\int^{T}_0 (-t_1-t_2 e^{-ik})e^{-i\omega t+2\cos{\omega t}} e^{i(m-m')\omega t} dt\\
&=(-t_1-t_2 e^{-ik})\sum_m \mathcal{J}_{m'+1-m}(-2i) (i)^{(m'+1-m)}\\
&=(-t_1-t_2 e^{-ik})\sum_m (-1)^{m'+1-m} \mathcal{I}_{(m'+1-m)}(-2)\\
\hat{H}^{off-diagonal}_{2,1}&=\frac{1}{T}\int^{T}_0 (-t_1-t_2 e^{ik})e^{i\omega t-2\cos{\omega t}} e^{i(m-m')\omega t} dt\\
&=(-t_1-t_2 e^{ik})\sum_m \mathcal{J}_{m'-1-m}(2i) (i)^{(m'-1-m)}\\
&=(-t_1-t_2 e^{ik})\sum_m (-1)^{m'-1-m} \mathcal{I}_{(m'-1-m)}(2)
\end{aligned}
\end{eqnarray}
where $\mathcal{J}_{\alpha}(x)$ and $\mathcal{I}_{\alpha}(x)$ are the Bessel and modified Bessel functions.  Through subequations (22-23), it is presented that the frequency relevant off-diagonal term contributes to  $\lim_{\omega\rightarrow\infty}\frac{1}{\omega}|\tilde{h}_{m-m'}|\neq0$ in high frequency limit, thus Wannier-Stark localization can not take place.
To restore the Wannier-Stark localization and predict topological phases from frequency space Floquet Hamiltonian, time dependent non-unitary deformation  is applied. To be specific, we take $\Gamma(t)=\bigoplus^{i=L}_{i=1} \left( \begin{smallmatrix}
             e^{-i\frac{\omega t}{2}+\cos{\omega t}} & 0 \\ 0 & e^{i\frac{\omega t}{2}-\cos{\omega t}}
          \end{smallmatrix}  \right) $, through subequation (15), it can be obtained the evolving equations
corresponding to the deformed Floquet states are $H'(t)=H'_0+H'_t$:
\begin{eqnarray}
H'_{0}=\left( \begin{smallmatrix}  0& -t_1-t_2 e^{-ik} \\ -t_1-t_2 e^{ik} & 0 \end{smallmatrix} \right)
\end{eqnarray}
\begin{eqnarray}
H'_{t}=\left( \begin{smallmatrix} (2p\cos{k}-\mu_0)\sin{\omega t} & 0 \\ 0 & -(2p\cos{k}-\mu_0)\sin{\omega t} \end{smallmatrix} \right)
\end{eqnarray}
The off-diagonal terms of the deformed Floquet Hamiltonian is:
\begin{eqnarray}
\begin{aligned}
\hat{H'}^{off-diagonal}&=\frac{1}{T}\int^{T}_0 \sum_m H'_t e^{i(m-m')\omega t} dt\\
\hat{H}^{off-diagonal}_{1,1}&=-\hat{H}^{off-diagonal}_{3,3}=\frac{1}{T}\int^{T}_0 (2p\cos{k}-\mu_0)\sin{\omega t} e^{i(m-m')\omega t} dt\\
&=\sum_m (-ip\cos{k}+\frac{i\mu_0}{2})[\delta_{m=m'-1}-\delta_{m=m'+1}]
\end{aligned}
\end{eqnarray}
From the form of subequation (24-26), it can be identified  $\frac{1}{\omega} (|\tilde{h'}_{n-n'}|)\approx 0$. Wannier-Stark localization is reobtained, and different Floquet topological phases can be predicted from the generation of Floquet band gaps.